\documentclass[prd,aps,superscriptaddress,twocolumn,floatfix,nofootinbib]{revtex4}
\pdfoutput=1

\usepackage{amsfonts}
\usepackage{amsmath}
\usepackage{amssymb}
\usepackage{bm}
\usepackage{dcolumn}
\usepackage{graphicx}   
\usepackage[latin1]{inputenc}
\usepackage{latexsym}
\usepackage{rotating}
\usepackage{hyperref}
\usepackage{graphicx}
\usepackage{color}

\newcommand\be{\begin{equation}}
\newcommand\ba{\begin{eqnarray}}
\newcommand\ee{\end{equation}}
\newcommand\ea{\end{eqnarray}}

\begin{document}

\title{Unitarity Problems for an Effective Field Theory Description of Early Universe
Cosmology}

\author{Robert Brandenberger}
\email{rhb@physics.mcgill.ca}
\affiliation{Department of Physics, McGill University, Montr\'{e}al,
  QC, H3A 2T8, Canada}

\author{Vahid Kamali}
\email{vkamali@ipm.ir}
\affiliation{Department of Physics, McGill University, Montr\'{e}al,
  QC, H3A 2T8, Canada}
\affiliation{
  Department of Physics, Bu-Ali Sina (Avicenna) University, Hamedan 65178,
  016016, Iran}
\affiliation{
  School of Physics, Insitute for Research in Fundamental Sciences (IPM),
  19538-33511, Tehran, Iran}

%\date{\today}

%%%%%%%%%%%%%%%%%%%%%%%%%%%%%%%%%%%%%%%%%%%%%%%%%%%%%%%%%%%%%%%%%
\begin{abstract}

In the context of Effective Field Theory, the Hilbert space of states increases in an expanding universe. Hence, the time evolution cannot be unitary. The formation of structure is usually studied using effective field theory techniques. We study the constraints on effective field theory analyses of early universe models which come from demanding that the factor of the space of states corresponding to length scales where the primordial fluctuations are manifest does not suffer from the unitarity problem.  For bouncing and emergent cosmologies, no constraints arise provided that the energy scale of the bounce or emergent phases is smaller than the ultraviolet (UV) cutoff scale. On the other hand, in the case of the inflationary scenario, non-trivial upper bounds on the energy scale of inflation arise.

\end{abstract}
%%%%%%%%%%%%%%%%%%%%%%%%%%%%%%%%%%%%%%%%%%%%%%%%%%%%%%%%%%%%%%%%%

%\pacs{98.80.Cq}
\maketitle

%%%%%%%%%%%%%%%%%%%%%%%%%%%%%%%%%%%%%%%%%%%%%%%%%%%%%%%%%%%%%%%%%
\section{Introduction} 
\label{sec:intro}

Most early universe models are based on an effective field theory analysis. In this framework, matter fields and metric fluctuations are quantized on a homogeneous and isotropic cosmological background, and the initial conditions for the fluctuatios are often taken to be quantum vacuum perturbations \cite{Mukh, Starob2}. Examples are slow-roll inflation models \cite{Guth, Linde} and the Ekpyrotic bouncing cosmology \cite{Ekp}. In both cases, a scalar matter field is introduced to yield the cosmological background evolution desired, and it is initial vacuum fluctuations of this scalar field (coupled to the induced metric fluctuations) which are responsible for the origin of structure in the Universe.

In an effective field theory analysis, the fields are expanded in Fourier modes, and each Fourier mode is quantized as a harmonic oscillator. Note that the Fourier modes have fixed comoving wavenumber $k$, and hence their corresponding physical wavelength increases as the Universe expands. To avoid an ultraviolet (UV) catastrophy, a cutoff $k_{phys} < k_{\Lambda}$ on the physical wavenumber is required. Note that the UV has to correspond to a fixed physical scale since it is determined by local physics. An upper bound on the cutoff scale $k_{\Lambda}$ is the Planck energy $m_{pl}$. As pointed out a long time ago in \cite{Weiss}, this leads to a serious problem for any effective field theory treatment in an expanding cosmology: to maintain a fixed physical scale for the UV cutoff there needs to be a continuous production of modes. It appears that the Hilbert space is time-dependent. As recently discussed in \cite{Andy}, this implies that the time evolution cannot be unitary and is at most isometric.

In this note we will explore the consequences of this problem for effective field theory treatments of various early universe scenarios. If an effective field theory model of the early universe scenario is to be predictive, then scales with $k < k_c$ on which fluctuations are thought to be of primordial origin must have a physical wavelength which remains larger than the cutoff length scale at all times. We will explore the consequences of this demand for inflationary models as well as for a class of bouncing and emergent scenarios \footnote{See e.g. \cite{RHBalt} for reviews of alternatives to inflationary cosmology.}.

The unitarity problem for effective field theory models of inflation was discussed a number of years ago under the name {\it Trans-Planckian Problem} (TPP) for cosmological perturbations \cite{Jerome} where it was pointed out that, if the wavelength corresponding to the mode whose current wavelength is equal to the current Hubble radius \footnote{We remind the reader that the Hubble radius is $l_H = H^{-1}$, where $H$ is the expansion rate. On sub-Hubble scales, fluctuations oscillate while they are frozn in on super-Hubble scales (see e.g. \cite{MFB, RHBrev} for reviews of the theory of cosmological perturbations). In Standard Big Bang cosmology, the Hubble radius equals (modulo a numerical factor of order one) the causal horizon, while in any early universe model which can explain the near isotropy of the Cosmic Microwave Background (CMB), the horizon must be much larger than the Hubble radius at late times.} is smaller than the Planck length at the beginning of inflation, then new physics must enter into the determination of the spectrum of fluctuations on all measurable scales. In a body of works (see e.g. \cite{Easther, Danielsson, Niemeyer, Bozza, Jackson}) it was proposed to set initial conditions on a time-like ``new physics hypersurface'' where each $k$ mode is initiated when its physical wavelength equals the cutoff length, and the influence of various choices of the initial conditions on the final spectrum of cosmological perturbations was explored (see \cite{JeromeRev} for a review of this and other approaches). Evidently, this prescription does not correspond to unitary evolution. 

Recently, Bedroya and Vafa \cite{Bedroya1} formulated the {\it Trans-Planckian Censorship Conjecture} (TCC) which states that in no consistent quantum theory of gravity the situation will arise that some mode with initial wavelength smaller than the Planck length becomes super-Hubble. As discussed in \cite{RHBrev2}, this condition corresponds to demanding that an observer measuring only super-Hubble modes is shielded (at least at the level of a linear analysis) from the non-unitarity of the effective field theory model in the expanding universe. The implications of the TCC for inflationary cosmology was discussed in \cite{Bedroya2} where it was found that, assuming almost exponential expansion of space during the phase of inflation, and taking reheating to be rapid, an restrictive upper bound $\eta < 3 \times 10^9 {\rm{GeV}}$ on the allowed energy scale $\eta$ of inflation results. Allowing for power law inflation and/or delayed reheating leads to a weakening of the bound \cite{TCC1}, while taking into account the pre-inflationary phase leads to a strengthening \cite{TCC2}.

The unitarity problem \cite{Weiss, Andy} for an effective field theory description of an expanding universe is closely related both to the {\it Trans-Planckian Problem} (TPP) and the {\it Trans-Planckian Censorship Conjecture} (TCC). If we demand that the factor of the state corresponding to all modes which crossed the Hubble radius during the primodial phase is unaffected by the non-unitarity of the time evolution \footnote{Note that it is only super-Hubble modes which are squeezed and can classicalize.}, our criterion becomes identical to the TCC, while if we only demand that modes which are currently of Hubble length and larger do not suffer from the unitarity problem, then we obtain the conclusions of the TPP \cite{Jerome}. It is natural to demand that the factor of the Hilbert space for which there is observational evidence that their origin is primordial is not affected by the unitarity issues. We will denote by $k_c$ the largest wavenumber for which this is the case, and we will study the dependence of the constraints on $k_c$. A conservative choice for $k_c$ is the value corresponding to the smallest angular scale which can be reliably measured in the cosmic microwave background.
 
Most of the following discussion will focus on the inflationary scenario, the one which is most affected by the TCC and the unitarity problems. We will also briefly mention the implications for alternatives to inflation such as bouncing and emergent cosmologies \footnote{See e.g. \cite{Alt} for a review of some alternatives to inflation.}. Note that our discussion only affects models which are based on effective field theory techniques. There have recently been attempts to construct phases of accelerated cosmological expansion as coherent states (coherent states of gravitons \cite{Dvali} or Glauber-Sudarshan coherent states constructed from superstring theory \cite{Keshav}). These scenarios are unaffected by our constraints. 

Note that our analysis is independent of what the ultraviolet completion of the cosmological model might be. If we assume that the underlying ultraviolet complete theory is superstring theory, then it is known that effective field theories consistent with the ultraviolet completion are severly constrained - this is the {\it swampland program} (see \cite{original} for original references and \cite{swamprevs} for recent reviews).
  
In the following section we will explain our unitarity considerations. Then, in Section \ref{secttwo} we will study the implications of our unitarity criterion for a canonical inflation model with (nearly) constant Hubble expansion rate, and with rapid transition to the radiation phase of Standard Cosmology at the end of inflation, i.e. assuming that the period during which the energy is transferred from the inflaton to radiation lasts for less than one Hubble expansion time. In Section \ref{sectthree} we generalize the considerations to power law inflation models and models with delayed reheating, and in Section \ref{sectfour} we briefly discuss the consequences for bouncing and emergent scenarios. We assume standard {}Friedmann-Robertson-Walker-Lemaitre cosmology with the space-time metric given by 
\be ds^2 = dt^2 - a(t)^2 d{\bf x}^2 , \ee
where $t$ is physical time, ${\bf x}$ are the comoving spatial coordinates and $a(t)$ is the scale factor. The expansion rate is $H(t) \equiv {\dot{a}} / a$, the dot indicating a derivative with respect to time. We work in natural units in which Planck's constant, the speed of light and Boltzmann's constant are set to $1$. The Planck mass $m_{pl}$ is given by $G = m_{pl}^{-2}$, where $G$ is Newton's gravitational constant. Temperatures are denoted by $T$, and cosmological redshift by $z$. Relevant times are the current time $t_0$, the time $t_{rec}$ of recombination, and the time $t_{eq}$ of equal matter and radiation.
\\

%%%%%%%%%%%%%%%%%%%%%%%%%%%%%%%%%%%%%%
\section{Unitarity Considerations}
\label{sectone}

In an effective field theory we canonically quantize each Fourier mode of our fields. The modes have constant comoving length and are labelled by their comoving wavenumber $k$. We can write the total Hilbert space as
\be
{\cal{H}} \, = \, \otimes_k {\cal{H}}_k \, ,
\ee
where ${\cal{H}}_k$ is the harmonic oscillator Hilbert space of mode $k$. To avoid singularities and to remain in the range of applicability of an effective field theory, there needs to be an ultraviolet cutoff scale $k_{\Lambda}$. Note that this is a physical cutoff scale, not a comoving one.

The unitarity problem for an effective field theory analysis comes from the fact that, in order to maintain the ultraviolet cutoff at a fixed physical scales, we must continuously add modes to the Hilbert space \cite{Weiss, Andy}. Thus, if $k_i$ denotes the comoving mode whose physical wavenumber equals the cutoff scale at some initial time $t_i$ and which hence determines the initial Hilbert space ${\cal{H}}_i$, then at a later time $t$ the Hilbert space ${\cal{H}}(t)$ is larger than the initial Hilbert space:
\be \label{factoring}
{\cal{H}}(t) \, = \, \bigl( \otimes_{k > k_i} {\cal{H}}_k \bigr) \otimes {\cal{H}}_i \, .
\ee
 A state $|\psi>_i$ in the initial Hilbert space can be isometrically evolved \cite{Andy} into the following product state in the full Hilbert space at a later time
 \be
 |\psi>_i \, \rightarrow \, |{\tilde{\psi}}(t)> |\psi_i> \, ,
 \ee
 where
 \be
 |{\tilde{\psi}}(t)> \,\,  \in \,\,  \otimes_{k > k_i} {\cal{H}}_k \, 
 \ee
 is not determined by the effective field theory, but can be chosen such it has unit norm, in which case the norm of $|\psi>(t)$ in ${\cal{H}}(t)$ equals the form of $|\psi>_i$ in the initial Hilbert space. It is in this sense that the evolution is isometric. But note that effective field theory cannot tell use about the nature of  $|{\tilde{\psi}}(t)>$ except for its normalization. This is the {\it Trans-Planckian Problem} for effective field theories in an expanding universe \cite{Jerome}.
 
What does this imply for models to explain the origin of the observed structure in the universe based on some early universe model? Today, we observe the state of inhomogeneities, e.g. the state of the matter distribution and the state of CMB anisotropies. On large scales, the fluctuations are in the linear regime, and hence they have evolved linearly from early times to the present time. On shorter scales nonlinearities play an important role, and there is vigorous mode mixing. Power on large scales can flow down to smaller scales. 

It is clear that we cannot evolve any state in the current effective field theory Hilbert space ${\cal{H}}(t_0)$ into the past without hitting the unitarity wall. The mode whose physical wavelength today is the Planck length has just now been created. But this mode obviously has no influence on any observables. Which modes should we then be able to evolve safely backwards in time? One answer is given by the TCC \cite{Bedroya1, RHBrev}: we demand that all modes which have exited the Hubble radius at some early time (and which have thus been squeezed and classicalized) have evolved safely. But once again, at the high $k$ end such modes are far removed from what can be observed today. We propose a conservative approach in which we pick a scale $k_c$ above which (i.e. for wavelength larger than $k_c^{-1}$) we demand that the evolution backwards in time does not hit the unitarity wall.  We will keep the value of $k_c$ as a free parameter. It is clear that this value is much larger than the value of $k$ corresponding to the first acoustic peak of the angular power spectrum of CMB anisotropies (which corresponds to the length of the Hubble radius at the time $t_{rec}$ of recombination and which we denote by $k_1$). The extreme conservative choice is to take $k_c$ to correspond to the largest angular harmonic $l$ value for which CMB experiments have explored the primordial anisotropy spectrum. 

Based on the above considerations, we will impose the following criterion: We demand that the mode corresponding to $k_c$ starts out at some initial time $t_b$ with a wavelength larger than the Planck length (which we take to be the ultraviolet cutoff length), i.e. 
\be \label{criterion}
l_{phys}(t_b) (k_c) \, > \, m_{pl}^{-1} \, .
\ee

%%%%%%%%%%%%%%%%%%%%%%%%%%%%%%%%%%%%%%
\section{Application to Canonical Inflation}
\label{secttwo} 

We will first apply the criterion (\ref{criterion}) to the case of canonical (allmost) exponential inflation. We will take the expansion rate $H$ during the inflationary phase to be constant. The number of e-foldings of inflation will be denoted by $N$, the time of the onset of inflation by $t_i$ and the end time by $t_R$. We will also assume that the transition between the end of inflation and the beginning of the radiation phase of Standard Big Bang cosmology takes place in less than one Hubble expansion time, which generically happens if the preheating instability \cite{TB} (see \cite{RHrevs} for reviews) which governs the transfer of energy between the inflaton field and Standard Model matter is efficient. The redshift corresponding to the end of inflation is
\be \label{endredshift}
z_R \, = \, \frac{\eta}{T_0} \, ,
\ee
where $\eta$ is the energy scale during inflation, and $T_0$ is the current CMB temperature. The Hubble expansion rate $H$ during inflation is related to $\eta$ via the Friedmann equation:
\be
H^2 \, = \, \frac{8 \pi}{3} m_{pl}^{-2} \eta^4 \, .
\ee

We will first demand that the criterion (\ref{criterion}) be satisfied at the beginning of inflation (and later consider modifications if the pre-inflationary evolution is taken into account). Thus, we set $t_b = t_i$. Tracing back the length scale corresponding to the mode $k_c$ to the beginning of inflation yields
\ba
l_{phys}(t_i) (k_c) \, &=& e^{-N} \frac{z_{rec}}{z_R} \frac{k_1}{k_c} t_{rec} \\
&=& e^{-N} z_R^{-1} z_{rec}^{-1/2} \frac{k_1}{k_c} t_0 \, . \nonumber
\ea
Using the Friedmann equation to substitute for $t_0$ in terms of the temperature $T_{rec}$ at recombination (and neglecting the dark energy component to today's energy density) we obtain
\be
l_{phys}(t_i) (k_c) \, = \, (6 \pi)^{-1/2} e^{-N} \frac{m_{pl}}{\eta T_{rec}} \frac{k_1}{k_c}
\bigl( \frac{z_{rec}}{z_{eq}}  \bigr)^{1/2} \, ,
\ee
which needs to be larger than $m_{pl}^{-1}$ by our criterion. This yields the following upper bound on the number of e-foldings of inflation
\be \label{upper}
e^N \frac{\eta}{m_{pl}} \, < \, (6 \pi)^{-1/2} \frac{m_{pl}}{T_{rec}} \frac{k_1}{k_c}
\bigl( \frac{z_{rec}}{z_{eq}}  \bigr)^{1/2} \, .
\ee

A lower bound on the duration of inflation results by demanding that the current Hubble radius emerges from inside the Hubble radius at the beginning of inflation \footnote{If this condition is not satisfied then inflation cannot explain the origin of the structure which is observed on cosmological scales.}. The resulting lower bound on the number of e-foldings $N$ is
\be \label{lower}
e^N \, > \, \frac{2}{3} z_{rec} z_{eq}^{-1/2} \frac{\eta}{T_{rec}} \, .
\ee
The upper and lower bounds (\ref{upper}) and (\ref{lower}) on the duration of inflation are consistent provided that
\be \label{res1}
\frac{\eta}{m_{pl}} \, < \, \bigl( \frac{1}{2} \bigr)^{1/2} \bigl( \frac{3}{ 2 \pi} \bigr)^{1/4} z_{rec}^{-1/4} 
\bigl( \frac{k_1}{k_c} \bigr)^{1/2} \, .
\ee

If we make the very conservative choice that only scales which are measured in the CMB are protected from the unitarity problem, then $k_c \sim 10^2 k_1$ and we find that canonical inflation with the value of $\eta$ at the scale of Grand Unification is viable. However, if we are less conservative and demand that the evolution backwards in time of scales of much larger $k$ is well behaved, then the energy scale $\eta$ begins to be seriously constrained from above. If we take $k_c$ to correspond to the scale which exits the Hubble radius at the end of inflation, we recover the tight constraint $\eta < 3 \times 10^{9} {\rm{GeV}}$ from the TCC \cite{Bedroya2}.

A condition which is stricter than (\ref{res1}) can be obtained by taking into account the evolution before the onset of inflation. Assuming that the evolution is adiabatic between the Planck time $t_{pl}$ (when the temperature is $m_{pl}$ and the onset of inflation, the criterion (\ref{criterion}) with $t_b = t_{pl}$ is stricter by a factor of $\eta / m_{pl}$ then if we use $tt_b = t_i$ corresponding to the beginning of inflation. Hence, the upper bound (\ref{upper}) on the duration of inflation becomes
\be
e^N  \, < \, (6 \pi)^{-1/2} \frac{m_{pl}}{T_{rec}} \frac{k_1}{k_c}
\bigl( \frac{z_{rec}}{z_{eq}}  \bigr)^{1/2} \, .
\ee
Combined with the lower bound (\ref{lower}) we get the condition
\be \label{res2}
\frac{\eta}{m_{pl}} \, < \,  \frac{1}{2}  \bigl( \frac{3}{ 2 \pi} \bigr)^{1/2} z_{rec}^{-1/2} 
\bigl( \frac{k_1}{k_c} \bigr) \, ,
\ee
which is a stricter bound on $\eta$ and already puts pressure on inflation at the Grand Unification scale if we take the very conservative choice of $k_c = 10^2 k_1$.

%%%%%%%%%%%%%%%%%%%%%%%%%%%%%%%%%%%%%%
\section{Generalization to Power-Law Inflation}
\label{sectthree}

There are two generalizations of the above analysis which we can consider. The first is the generalization to power law inflation, maintaining the assumption of rapid thermalization. In this case, the effective temperature $\eta_R$ at the end of the period of inflation is lower than the effective temperature $\eta_i$ at the beginning. The condition (\ref{upper}) remains the same, with $\eta_R$ replacing $\eta$
\be \label{upper2}
e^N \frac{\eta_R}{m_{pl}} \, < \, (6 \pi)^{-1/2} \frac{m_{pl}}{T_{rec}} \frac{k_1}{k_c}
\bigl( \frac{z_{rec}}{z_{eq}}  \bigr)^{1/2} \, ,
\ee
while the lower bound on the duration of inflation takes the form
\be \label{lower2}
e^N \, > \, \frac{2}{3} z_{rec} z_{eq}^{-1/2} \frac{\eta_i^2}{\eta_R T_{rec}} \, .
\ee
These two bounds are self-consistent provided that
\be
\frac{\eta_i}{m_{pl}} \, < \, \bigl( \frac{1}{2} \bigr)^{1/2} \bigl( \frac{3}{ 2 \pi} \bigr)^{1/4} z_{rec}^{-1/4} 
\bigl( \frac{k_1}{k_c} \bigr)^{1/2} \, ,
\ee
which is the same as (\ref{res1}) with $\eta_i$ in place of $\eta$. The result (\ref{res2}) generalizes in the same way to this case, i.e. by $\eta_i$ replacing $\eta$.

The second generalization is to consider delayed reheating. Specifically, let us assume that after the end of inflation until a time $t_{rh}$, the universe is not dominated by radiation, but by a fluid (denoted w-fluid) with an equation of state parameter $w$ (e.g. $w = 0$ for a matter phase which could occur during the time when the inflaton field is coherently oscillating about the minimum of its potential). In this case, the redshift ${\tilde{z}}_R$ at the end of inflation is not given by $z_R$ (see (\ref{endredshift})), but rather by
\be
{\tilde{z}}_R \, = \, z_R \bigl( \frac{\eta_R}{T_{rh}} \bigr)^E \, ,
\ee
where the exponent $E$ is given by
\be
E \, = \, \frac{4}{3(1 + w)} - 1 \, ,
\ee
where $T_{rh}$ is the temperature at the beginning of the radiation phase. To obtain this result, we have assumed that all of the energy in the w-fluid is transformed to radiation at the time $t_{rh}$, yielding a radiation temperature $T_{rh}$. 

From the analysis of the previous section it is clear that this change effects the upper and lower bounds on the duration of inflation in the same way, and hence does not change the resulting conditions on the energy scale of inflation. It does, however, effect the individual constraints on $e^{N}$ by a factor of ${\tilde{z}}_R / z_R$.

%%%%%%%%%%%%%%%%%%%%%%%%%%%%%%%%%%%%%%
\section{Implications for Bouncing and Emergent Cosmologies}
\label{sectfour}

\subsection{Bouncing Cosmologies}

Bouncing and emergent scenarios can provide alternatives to cosmological inflation for explaining the solving the problems of Standard Big Bang cosmology and for providing a causal mechanism for explaining the origin of the observed structure in the universe (see e.g. \cite{RHBalt} for reviews). Among bouncing scenarios, the Ekpyrotic scenario \cite{Ekp} has a particular appeal. It assumes a very slow phase of contraction which can be obtained in the context of an effective field theory analysis by coupling matter with an equation of state $w \gg 1$ to the Einstein-Hilbert gravitational action. Scalar fields with a negative exponential potentials can yield such matter, and negative exponential potentials for scalar fields are also well motivated from superstring theory. This scenario has the feature that anisotropies and spatial curvature are diluted in the contracting phase \cite{Erickson}, and that the homogeneous background trajectory is a global attractor in initial condition space \cite{Ijjas}. The main challenges for Ekpyrotic cosmology are to obtain a well-controlled transition from Ekpyrotic contraction to Standard Big Bang expansion, and to obtain a roughly scale-invariant spectrum of cosmological perturbations starting with quantum vacuum perturbations in the early phase of contraction. Both of these challenges can be solved \cite{Ziwei} by postulating that the bounce is mediated by an S-Brane \cite{Sbrane}, a space-filling relativistic term (motivated by string theory) which arises in a low energy effective action at the time when the energy density reaches the string scale (denoted here by $\eta$) and towers of new states become low mass. Such a brane has vanishing energy density and negative pressure (because is has tension) and hence violates the weak and null energy weak energy conditions, and it thus can mediate the required transition between the contracting and expanding phases \footnote{See \cite{NewEkp} for other ways of obtaining a bounce.}. Making use of the matching conditions for cosmological fluctuations discussed in \cite{Durrer}, it can be shown that initial vacuum spectra early in the contracting phase are transformed into nearly scale-invariant spectra of both curvature fluctuations and gravitational waves after the bounce \cite{Ziwei}. After the bounce, the equation of state of matter is that of radiation. 

Since in the post-bounce phase the comoving Hubble radius is contracting, no modes ever exit the Hubble radius. Hence, the TCC is trivially satsfied from this point of view. During the contracting phase, modes do exit the Hubble radius, but as long as the bounce energy scale is smaller than the Planck scale, no modes exit the Hubble radius which at any time had a wavelength smaller than the Planck length. Thus, the TCC is satisfied. Our unitarity constraint for the evolution of cosmological perturbations is also satisfied as long as we focus on the set of modes which at the bounce time had a wavelength larger than the Planck length. Their wavelength $\lambda$ today is
\be
\lambda \, > \, m_{pl}^{-1} \frac{\eta}{T_0} \, .
\ee

Another way to see that the bouncing scenario is basically safe against the uncertainties coming from the breakdown of unitarity is to follow the analysis of Section \ref{secttwo} and trace back to $T = \eta$ the wavelength of the smallest mode whose fluctuations we take to be primordial, and to demand
\be \label{bouncecond}
l_{\rm{phys}}(T = \eta) \, > \, m_{pl}^{-1} \, .
\ee
The physical length $l_{\rm{phys}}$ at the bounce time can be computed easilly since the expanding phase is given by Standard Big Bang cosmology. Using the same notation as in Section \ref{secttwo} we obtain
\be \label{value}
l_{\rm{phys}}(T = \eta) \, = (6 \pi)^{-1/2} \bigl( \frac{k_1}{k_c} \bigr) 
\bigl( \frac{z_{rec}}{z_{eq}} \bigr)^{1/2} \frac{m_{pl}}{T_{rec}} \, ,
\ee
such that (\ref{bouncecond}) is trivially satisfied as long as the bounce energy scale $\eta$ is smaller than the Planck mass.

\subsection{Emergent Cosmologies}

Emergent cosmologies yield another class of alternatives to the inflationary scenario. Here, it is postulated that the radiation phase of Standard Big Bang cosmology is preceded by an emergent phase. If the perturbations in the emergent phase are thermal fluctuations with holographic scaling of thermal correlation functions, it follows that the induced curvature perturbations and gravitational waves will be scale invariant \cite{NBV}. One example of an emergent scenario is {\it String Gas Cosmology} (SGC) \cite{BV} where it is assumed that the emergent phase is a quasi-static gas of strings at a temperature close to the Hagedorn temperature \cite{Hagedorn}, the maximal temperature of a gas of closed strings. As discussed in \cite{BV}, string winding modes can annihilate into string loops only in three large spatial dimensions. The phase transition between the emergent phase and the radiation phase of Standard cosmology is given by the annihilation of the strings winding our three spatial cycles. The power spectrum of the induced cosmological fluctuations is proportional to $(\eta / m_{pl})^4$, and hence if the string scale is about $\eta \sim 10^{17} {\rm{GeV}}$, the observed amplitude of the power spectrum is obtained. Interestingly, this value for the string scale is of the same order as the scale preferred from the point of view of heterotic string phenomenology (see \cite{GSW} for a review). More recent examples of emergent cosmologies are the matrix model cosmology scenario of \cite{Brahma} or the model of \cite{Vafa} where the emergent phase is a topological phase.

Since the evolution of an emergent cosmology after the phase transition is the same as in Standard Big Bang cosmology, and the same as the post-bounce evolution in a bouncing scenario, the results (\ref{bouncecond}) and (\ref{value}) apply, and the models are trivially safe from the unitarity constraints discussed in this paper.

%%%%%%%%%%%%%%%%%%%%%%%%%%%%%%%%%%%%%%%%%%%%%%%%%%%%%%%%%%%%%%%%%
\section{Conclusions} 
\label{conclusion}

 We have studied the unitarity problem for effective field theory descriptions of the early universe. Demanding that the part of the Hilbert space corresponding to fluctuation scales which we have good reason to believe to be of primordial origin were always in the effective field theory Hilbert space leads to constraints. In the case of bouncing and emergent cosmologies, an effective field theory description is safe as long as the energy scale of the bounce or of the emergent phase is smaller than the UV cutoff scale. The energy scale of nflationary models, on the other hand, is more tightly constrained: the upper bound on the energy scale of inflation is given by (\ref{res1}) and is hence parametrically smaller than the UV cutoff scale (which, here, is taken to be the Planck scale). Taking into account a pre-inflationary radiation phase leads to a tighter constraint, namely (\ref{res2}). If we demand that all modes which were ever super-Hubble were in the initial effective field theory Hilbert space, our results reduce to those obtained using the Trans-Planckian Censorship Conjecture \cite{Bedroya1, Bedroya2}.

%%%%%%%%%%%%%%%%%%%%%%%%%%%%%%%%%%%%%%%%%%%%%%%%%%%%%%%%%%%%%%%%%
\section*{Acknowledgement}

\noindent 
 The research at McGill is supported in
part by funds from NSERC and from the Canada Research Chair
program. RB is grateful for hospitality of the Institute for
Theoretical Physics and the Institute for Particle Physics and
Astrophysics of the ETH Zurich during the completion of this project.
VK would like to acknowledge the McGill University
Physics Department for hospitality and partial financial support.  

%%%%%%%%%%%%%%%%%%%%%%%%%%%%%%%%%%%%%%%%%%%%%%%%%%%%%%%%%%%%%%%%%


\begin{thebibliography}{99}

\bibitem{Mukh}
V. Mukhanov and G. Chibisov,
 ``Quantum Fluctuation And Nonsingular Universe. (In Russian),''
 JETP Lett.\  {\bf 33}, 532 (1981) [Pisma Zh.\ Eksp.\ Teor.\ Fiz.\  {\bf 33}, 549 (1981)].
 %%CITATION = ZFPRA,33,549;%%
 
 \bibitem{Starob2}
 A.~A.~Starobinsky,
``Spectrum of relict gravitational radiation and the early state of the universe,''
  JETP Lett.\  {\bf 30}, 682 (1979)
  [Pisma Zh.\ Eksp.\ Teor.\ Fiz.\  {\bf 30}, 719 (1979)].
  %%CITATION = JTPLA,30,682;%%
  
\bibitem{Guth}
A.~H.~Guth,
 ``The Inflationary Universe: A Possible Solution to the Horizon and Flatness Problems,''
 Phys.\ Rev.\ D {\bf 23}, 347 (1981)
 [Adv.\ Ser.\ Astrophys.\ Cosmol.\  {\bf 3}, 139 (1987)].
 doi:10.1103/PhysRevD.23.347;\\
 %%CITATION = doi:10.1103/PhysRevD.23.347;%%
 R.~Brout, F.~Englert and E.~Gunzig,
 ``The Creation Of The Universe As A Quantum Phenomenon,''
 Annals Phys.\  {\bf 115}, 78 (1978);\\
 %%CITATION = APNYA,115,78;%%
 A.~A.~Starobinsky,
 ``A New Type Of Isotropic Cosmological Models Without Singularity,''
 Phys.\ Lett.\ B {\bf 91}, 99 (1980);\\
 %%CITATION = PHLTA,B91,99;%%
 K.~Sato,
 ``First Order Phase Transition Of A Vacuum And Expansion Of The Universe,''
 Mon.\ Not.\ Roy.\ Astron.\ Soc.\  {\bf 195}, 467 (1981);\\
 %%CITATION = MNRAA,195,467;%%
L.~Z.~Fang,
  ``Entropy Generation in the Early Universe by Dissipative Processes Near the Higgs' Phase Transitions,''
  Phys.\ Lett.\  {\bf 95B}, 154 (1980).
  doi:10.1016/0370-2693(80)90421-9.
  %%CITATION = doi:10.1016/0370-2693(80)90421-9;%%
  
\bibitem{Linde} 
  A.~D.~Linde,
  ``A New Inflationary Universe Scenario: A Possible Solution of the Horizon, Flatness, Homogeneity, Isotropy and Primordial Monopole Problems,''
  Phys.\ Lett.\  {\bf 108B}, 389 (1982)
  [Adv.\ Ser.\ Astrophys.\ Cosmol.\  {\bf 3}, 149 (1987)].
  doi:10.1016/0370-2693(82)91219-9;\\
  %%CITATION = doi:10.1016/0370-2693(82)91219-9;%%
A.~Albrecht and P.~J.~Steinhardt,
  ``Cosmology for Grand Unified Theories with Radiatively Induced Symmetry Breaking,''
  Phys.\ Rev.\ Lett.\  {\bf 48}, 1220 (1982)
  [Adv.\ Ser.\ Astrophys.\ Cosmol.\  {\bf 3}, 158 (1987)].
  doi:10.1103/PhysRevLett.48.1220.
  %%CITATION = doi:10.1103/PhysRevLett.48.1220;%%

\bibitem{Ekp}
  J.~Khoury, B.~A.~Ovrut, P.~J.~Steinhardt and N.~Turok,
 ``The Ekpyrotic universe: Colliding branes and the origin of the hot big
 bang,''
 Phys.\ Rev.\ D {\bf 64}, 123522 (2001) [hep-th/0103239];\\
 %%CITATION = HEP-TH/0103239;%%
J.~Khoury, B.~A.~Ovrut, N.~Seiberg, P.~J.~Steinhardt and N.~Turok,
  ``From big crunch to big bang,''
  Phys.\ Rev.\ D {\bf 65}, 086007 (2002)
  doi:10.1103/PhysRevD.65.086007
  [hep-th/0108187].
  %%CITATION = doi:10.1103/PhysRevD.65.086007;%%
  
\bibitem{Weiss}
N.~Weiss,
  ``Constraints on Hamiltonian Lattice Formulations of Field Theories in an Expanding Universe,''
  Phys.\ Rev.\ D {\bf 32}, 3228 (1985).
  doi:10.1103/PhysRevD.32.3228
  %%CITATION = doi:10.1103/PhysRevD.32.3228;%%
  
\bibitem{Andy}
%\cite{Cotler:2022weg}
J.~Cotler and A.~Strominger,
``The Universe as a Quantum Encoder,''
[arXiv:2201.11658 [hep-th]].

\bibitem{RHBalt}
R.~H.~Brandenberger,
 ``Alternatives to the inflationary paradigm of structure formation,''
 Int.\ J.\ Mod.\ Phys.\ Conf.\ Ser.\  {\bf 01}, 67 (2011)
 doi:10.1142/S2010194511000109
 [arXiv:0902.4731 [hep-th]];\\
 %%CITATION = doi:10.1142/S2010194511000109;%%
 %\cite{Brandenberger:2010dk}
R.~H.~Brandenberger,
``Cosmology of the Very Early Universe,''
AIP Conf. Proc. \textbf{1268}, 3-70 (2010)
doi:10.1063/1.3483879
[arXiv:1003.1745 [hep-th]].

\bibitem{Jerome}
%\cite{Martin:2000xs}
J.~Martin and R.~H.~Brandenberger,
``The TransPlanckian problem of inflationary cosmology,''
Phys. Rev. D \textbf{63}, 123501 (2001)
doi:10.1103/PhysRevD.63.123501
[arXiv:hep-th/0005209 [hep-th]];\\
%\cite{Brandenberger:2000wr}
R.~H.~Brandenberger and J.~Martin,
``The Robustness of inflation to changes in superPlanck scale physics,''
Mod. Phys. Lett. A \textbf{16}, 999-1006 (2001)
doi:10.1142/S0217732301004170
[arXiv:astro-ph/0005432 [astro-ph]].

\bibitem{MFB}
V.F. Mukhanov, H.A. Feldman and R.H. Brandenberger,
 ``Theory of Cosmological Perturbations''
 Physics Reports \textbf{215}, 203 (1992).
 %%CITATION = doi:10.1016/0370-1573(92)90044-Z;%%
 
 \bibitem{RHBrev}
 R.~H.~Brandenberger,
 ``Lectures on the theory of cosmological perturbations,''
 Lect.\ Notes Phys.\  {\bf 646}, 127 (2004)
 doi:10.1007/978-3-540-40918-25
 [hep-th/0306071].
 %%CITATION = doi:10.1007/978-3-540-40918-2_5;%%

\bibitem{Easther}
%\cite{Easther:2001fi}
R.~Easther, B.~R.~Greene, W.~H.~Kinney and G.~Shiu,
``Inflation as a probe of short distance physics,''
Phys. Rev. D \textbf{64}, 103502 (2001)
doi:10.1103/PhysRevD.64.103502
[arXiv:hep-th/0104102 [hep-th]];\\
%\cite{Easther:2001fz}
R.~Easther, B.~R.~Greene, W.~H.~Kinney and G.~Shiu,
``Imprints of short distance physics on inflationary cosmology,''
Phys. Rev. D \textbf{67}, 063508 (2003)
doi:10.1103/PhysRevD.67.063508
[arXiv:hep-th/0110226 [hep-th]];\\
%\cite{Easther:2002xe}
R.~Easther, B.~R.~Greene, W.~H.~Kinney and G.~Shiu,
``A Generic estimate of transPlanckian modifications to the primordial power spectrum in inflation,''
Phys. Rev. D \textbf{66}, 023518 (2002)
doi:10.1103/PhysRevD.66.023518
[arXiv:hep-th/0204129 [hep-th]].

\bibitem{Danielsson}
%\cite{Danielsson:2002kx}
U.~H.~Danielsson,
``A Note on inflation and transPlanckian physics,''
Phys. Rev. D \textbf{66}, 023511 (2002)
doi:10.1103/PhysRevD.66.023511
[arXiv:hep-th/0203198 [hep-th]];\\
%\cite{Danielsson:2002qh}
U.~H.~Danielsson,
``Inflation, holography, and the choice of vacuum in de Sitter space,''
JHEP \textbf{07}, 040 (2002)
doi:10.1088/1126-6708/2002/07/040
[arXiv:hep-th/0205227 [hep-th]].

\bibitem{Niemeyer}
%\cite{Niemeyer:2002kh}
J.~C.~Niemeyer, R.~Parentani and D.~Campo,
``Minimal modifications of the primordial power spectrum from an adiabatic short distance cutoff,''
Phys. Rev. D \textbf{66}, 083510 (2002)
doi:10.1103/PhysRevD.66.083510
[arXiv:hep-th/0206149 [hep-th]].

\bibitem{Bozza}
%\cite{Bozza:2003pr}
V.~Bozza, M.~Giovannini and G.~Veneziano,
``Cosmological perturbations from a new physics hypersurface,''
JCAP \textbf{05}, 001 (2003)
doi:10.1088/1475-7516/2003/05/001
[arXiv:hep-th/0302184 [hep-th]].

\bibitem{Jackson}
%\cite{Jackson:2010cw}
M.~G.~Jackson and K.~Schalm,
``Model Independent Signatures of New Physics in the Inflationary Power Spectrum,''
Phys. Rev. Lett. \textbf{108}, 111301 (2012)
doi:10.1103/PhysRevLett.108.111301
[arXiv:1007.0185 [hep-th]];\\
%\cite{Schalm:2004qk}
K.~Schalm, G.~Shiu and J.~P.~van der Schaar,
``Decoupling in an expanding universe: Boundary RG flow affects initial conditions for inflation,''
JHEP \textbf{04}, 076 (2004)
doi:10.1088/1126-6708/2004/04/076
[arXiv:hep-th/0401164 [hep-th]].

\bibitem{JeromeRev}
%\cite{Brandenberger:2012aj}
R.~H.~Brandenberger and J.~Martin,
``Trans-Planckian Issues for Inflationary Cosmology,''
Class. Quant. Grav. \textbf{30}, 113001 (2013)
doi:10.1088/0264-9381/30/11/113001
[arXiv:1211.6753 [astro-ph.CO]].

\bibitem{Bedroya1}
  A.~Bedroya and C.~Vafa,
  ``Trans-Planckian Censorship and the Swampland,''
  arXiv:1909.11063 [hep-th].
  %%CITATION = ARXIV:1909.11063;%%
  
\bibitem{RHBrev2}
 R.~Brandenberger,
  ``Fundamental Physics, the Swampland of Effective Field Theory and Early Universe Cosmology,''
  arXiv:1911.06058 [hep-th];\\
  %%CITATION = ARXIV:1911.06058;%% 
  R.~Brandenberger,
  ``Trans-Planckian Censorship Conjecture and Early Universe Cosmology,''
  arXiv:2102.09641 [hep-th];\\
  %%CITATION = ARXIV:2102.09641;%%  
%\cite{Brandenberger:2021zib}
R.~Brandenberger,
``String Cosmology and the Breakdown of Local Effective Field Theory,''
[arXiv:2112.04082 [hep-th]].  
  
\bibitem{Bedroya2}
A.~Bedroya, R.~Brandenberger, M.~Loverde and C.~Vafa,
  ``Trans-Planckian Censorship and Inflationary Cosmology,''
Phys.\ Rev.\ D {\bf 101}, no. 10, 103502 (2020)
  doi:10.1103/PhysRevD.101.103502
  [arXiv:1909.11106 [hep-th]].
  %%CITATION = doi:10.1103/PhysRevD.101.103502;%%
  
\bibitem{TCC1}
  S.~Mizuno, S.~Mukohyama, S.~Pi and Y.~L.~Zhang,
  ``Universal Upper Bound on the Inflationary Energy Scale from the Trans-Planckian Censorship Conjecture,''
Phys.\ Rev.\ D {\bf 102}, no. 2, 021301 (2020)
  doi:10.1103/PhysRevD.102.021301
  [arXiv:1910.02979 [astro-ph.CO]];\\
  %%CITATION = doi:10.1103/PhysRevD.102.021301;%%
 M.~Dhuria and G.~Goswami,
  ``Trans-Planckian Censorship Conjecture and Non-thermal post-inflationary history,''
  Phys.\ Rev.\ D {\bf 100}, no. 12, 123518 (2019)
  doi:10.1103/PhysRevD.100.123518
  [arXiv:1910.06233 [astro-ph.CO]];\\
  %%CITATION = doi:10.1103/PhysRevD.100.123518;%%
  M.~Torabian,
  ``Non-Standard Cosmological Models and the trans-Planckian Censorship Conjecture,''
 Fortsch.\ Phys.\  {\bf 68}, no. 2, 1900092 (2020)
  doi:10.1002/prop.201900092
  [arXiv:1910.06867 [hep-th]];\\
  %%CITATION = doi:10.1002/prop.201900092;%%
 H.~H.~Li, G.~Ye, Y.~Cai and Y.~S.~Piao,
  ``Trans-Planckian censorship of multi-stage inflation and dark energy,''
 Phys.\ Rev.\ D {\bf 101}, no. 6, 063527 (2020)
  doi:10.1103/PhysRevD.101.063527
  [arXiv:1911.06148 [gr-qc]];\\
  %%CITATION = doi:10.1103/PhysRevD.101.063527;%%  
  V.~Kamali and R.~Brandenberger,
  ``Relaxing the TCC Bound on Inflationary Cosmology?,''
  Eur.\ Phys.\ J.\ C {\bf 80}, no. 4, 339 (2020)
  doi:10.1140/epjc/s10052-020-7908-8
  [arXiv:2001.00040 [hep-th]].
  %%CITATION = doi:10.1140/epjc/s10052-020-7908-8;%%
  
\bibitem{TCC2}
R. Brandenberger and E. Wilson-Ewing,
``Strengthening the TCC Bound on Inflationary Cosmology'',
JCAP {\bf 2003}, no. 03, 047 (2020)
  doi:10.1088/1475-7516/2020/03/047
  [arXiv:2001.00043 [hep-th]];\\
  %%CITATION = doi:10.1088/1475-7516/2020/03/047;%%
Y.~Cai and Y.~S.~Piao,
  ``Pre-inflation and Trans-Planckian Censorship,''
  Sci.\ China Phys.\ Mech.\ Astron.\  {\bf 63}, no. 11, 110411 (2020)
  doi:10.1007/s11433-020-1573-5
  [arXiv:1909.12719 [gr-qc]].
  %%CITATION = doi:10.1007/s11433-020-1573-5;%%
 
 \bibitem{Alt}
 R.~H.~Brandenberger,
  ``Introduction to Early Universe Cosmology,''
  PoS ICFI {\bf 2010}, 001 (2010)
  doi:10.22323/1.124.0001
  [arXiv:1103.2271 [astro-ph.CO]].
  %%CITATION = doi:10.22323/1.124.0001;%%
  
  \bibitem{Dvali}
G.~Dvali, C.~Gomez and S.~Zell,
  ``Quantum Break-Time of de Sitter,''
  JCAP {\bf 1706}, 028 (2017)
  doi:10.1088/1475-7516/2017/06/028
  [arXiv:1701.08776 [hep-th]];\\
  %%CITATION = doi:10.1088/1475-7516/2017/06/028;%%
  G.~Dvali and C.~Gomez,
  ``On Exclusion of Positive Cosmological Constant,''
  Fortsch.\ Phys.\  {\bf 67}, no. 1-2, 1800092 (2019)
  doi:10.1002/prop.201800092
  [arXiv:1806.10877 [hep-th]];\\
  %%CITATION = doi:10.1002/prop.201800092;%%
  G.~Dvali, C.~Gomez and S.~Zell,
  ``Quantum Breaking Bound on de Sitter and Swampland,''
  Fortsch.\ Phys.\  {\bf 67}, no. 1-2, 1800094 (2019)
  doi:10.1002/prop.201800094
  [arXiv:1810.11002 [hep-th]].
  %%CITATION = doi:10.1002/prop.201800094;%%
  
\bibitem{Keshav}
S.~Brahma, K.~Dasgupta and R.~Tatar,
  ``Four-dimensional de Sitter space is a Glauber-Sudarshan state in string theory,''
  arXiv:2007.00786 [hep-th];\\
  %%CITATION = ARXIV:2007.00786;%%                                                                                                                           
  S.~Brahma, K.~Dasgupta and R.~Tatar,
  ``de Sitter Space as a Glauber-Sudarshan State,''
  arXiv:2007.11611 [hep-th];\\
  %%CITATION = ARXIV:2007.11611;%%
  H.~Bernardo, S.~Brahma, K.~Dasgupta and R.~Tatar,
  ``Crisis on Infinite Earths: Short-lived de Sitter Vacua in the String Theory Landscape,''
  arXiv:2009.04504 [hep-th].
  %%CITATION = ARXIV:2009.04504;%%
 
\bibitem{original}
 H.~Ooguri and C.~Vafa, 
``On the Geometry of the String Landscape and the Swampland,'' 
  Nucl.\ Phys.\ B {\bf 766}, 21 (2007);\\
  %doi:10.1016/j.nuclphysb.2006.10.033 [hep-th/0605264].
    G.~Obied, H.~Ooguri, L.~Spodyneiko and C.~Vafa,
  ``De Sitter Space and the Swampland,''
  arXiv:1806.08362 [hep-th].
 %%CITATION = ARXIV:1806.08362;%%   
 
 \bibitem{swamprevs}
T.~D.~Brennan, F.~Carta and C.~Vafa,
 ``The String Landscape, the Swampland, and the Missing Corner,''
 PoS TASI {\bf 2017}, 015 (2017)
 doi:10.22323/1.305.0015
 [arXiv:1711.00864 [hep-th]];\\
 %%CITATION = doi:10.22323/1.305.0015;%%
 E.~Palti,
 ``The Swampland: Introduction and Review,''
 Fortsch.\ Phys.\  {\bf 67}, no. 6, 1900037 (2019)
  doi:10.1002/prop.201900037
  [arXiv:1903.06239 [hep-th]];\\
  %%CITATION = doi:10.1002/prop.201900037;%%    
 M.~van Beest, J.~Calderon-Infante, D.~Mirfendereski and I.~Valenzuela,
  ``Lectures on the Swampland Program in String Compactifications,''
  arXiv:2102.01111 [hep-th].
  %%CITATION = ARXIV:2102.01111;%%  

\bibitem{TB} 
J.~H.~Traschen and R.~H.~Brandenberger,
  ``Particle Production During Out-of-equilibrium Phase Transitions,''
  Phys.\ Rev.\ D {\bf 42}, 2491 (1990).
  doi:10.1103/PhysRevD.42.2491;\\
  %%CITATION = doi:10.1103/PhysRevD.42.2491;%%
 A.~D.~Dolgov and D.~P.~Kirilova,
  ``On Particle Creation By A Time Dependent Scalar Field,''
  Sov.\ J.\ Nucl.\ Phys.\  {\bf 51}, 172 (1990)
  [Yad.\ Fiz.\  {\bf 51}, 273 (1990)].
  %%CITATION = SJNCA,51,172;%%
  
 \bibitem{RHrevs}
  R.~Allahverdi, R.~Brandenberger, F.~Y.~Cyr-Racine and A.~Mazumdar,
  ``Reheating in Inflationary Cosmology: Theory and Applications,''
  Ann.\ Rev.\ Nucl.\ Part.\ Sci.\  {\bf 60}, 27 (2010)
  doi:10.1146/annurev.nucl.012809.104511
  [arXiv:1001.2600 [hep-th]];\\
  %%CITATION = doi:10.1146/annurev.nucl.012809.104511;%% 
M.~A.~Amin, M.~P.~Hertzberg, D.~I.~Kaiser and J.~Karouby,
  ``Nonperturbative Dynamics Of Reheating After Inflation: A Review,''
  Int.\ J.\ Mod.\ Phys.\ D {\bf 24}, 1530003 (2014)
  doi:10.1142/S0218271815300037
  [arXiv:1410.3808 [hep-ph]].
  %%CITATION = doi:10.1142/S0218271815300037;%% 
  
\bibitem{Erickson}
J.~K.~Erickson, D.~H.~Wesley, P.~J.~Steinhardt and N.~Turok,
  ``Kasner and mixmaster behavior in universes with equation of state w >= 1,''
  Phys.\ Rev.\ D {\bf 69}, 063514 (2004)
  doi:10.1103/PhysRevD.69.063514
  [hep-th/0312009].
  %%CITATION = doi:10.1103/PhysRevD.69.063514;%%

\bibitem{Ijjas}
%\cite{Ijjas:2016vtq}
A.~Ijjas and P.~J.~Steinhardt,
``Fully stable cosmological solutions with a non-singular classical bounce,''
Phys. Lett. B \textbf{764}, 289-294 (2017)
doi:10.1016/j.physletb.2016.11.047
[arXiv:1609.01253 [gr-qc]];\\
%\cite{Ijjas:2016tpn}
A.~Ijjas and P.~J.~Steinhardt,
``Classically stable nonsingular cosmological bounces,''
Phys. Rev. Lett. \textbf{117}, no.12, 121304 (2016)
doi:10.1103/PhysRevLett.117.121304
[arXiv:1606.08880 [gr-qc]].

\bibitem{Ziwei}
  R.~Brandenberger and Z.~Wang,
  ``Nonsingular Ekpyrotic Cosmology with a Nearly Scale-Invariant Spectrum of Cosmological Perturbations and Gravitational Waves,''
Phys.\ Rev.\ D {\bf 101}, no. 6, 063522 (2020)
  doi:10.1103/PhysRevD.101.063522
  [arXiv:2001.00638 [hep-th]];\\
  %%CITATION = doi:10.1103/PhysRevD.101.063522;%%
  R.~Brandenberger and Z.~Wang,
  ``Ekpyrotic cosmology with a zero-shear S-brane,''
  Phys.\ Rev.\ D {\bf 102}, no. 2, 023516 (2020)
  doi:10.1103/PhysRevD.102.023516
  [arXiv:2004.06437 [hep-th]];\\
  %%CITATION = doi:10.1103/PhysRevD.102.023516;%%
  R.~Brandenberger, K.~Dasgupta and Z.~Wang,
  ``Reheating after S-brane ekpyrosis,''
  Phys.\ Rev.\ D {\bf 102}, no. 6, 063514 (2020)
  doi:10.1103/PhysRevD.102.063514
  [arXiv:2007.01203 [hep-th]].
  %%CITATION = doi:10.1103/PhysRevD.102.063514;%%  

\bibitem{Sbrane}
 M.~Gutperle and A.~Strominger,
 ``Space - like branes,''
 JHEP {\bf 0204}, 018 (2002)
 doi:10.1088/1126-6708/2002/04/018
 [hep-th/0202210].
 %%CITATION = doi:10.1088/1126-6708/2002/04/018;%%

\bibitem{NewEkp}
 J.~L.~Lehners, P.~McFadden, N.~Turok and P.~J.~Steinhardt,
  ``Generating ekpyrotic curvature perturbations before the big bang,''
  Phys.\ Rev.\ D {\bf 76}, 103501 (2007) 
  doi:10.1103/PhysRevD.76.103501
  [hep-th/0702153 [HEP-TH]];\\ 
  %%CITATION = doi:10.1103/PhysRevD.76.103501;%%
E.~I.~Buchbinder, J.~Khoury and B.~A.~Ovrut, 
``New Ekpyrotic cosmology,'' Phys.\ Rev.\ D {\bf 76}, 123503 (2007)
doi:10.1103/PhysRevD.76.123503 [hep-th/0702154];\\ 
%%CITATION = doi:10.1103/PhysRevD.76.123503;%%
  P.~Creminelli and L.~Senatore, 
  ``A Smooth bouncing cosmology with scale invariant spectrum,'' 
  JCAP {\bf 0711}, 010 (2007)
  doi:10.1088/1475-7516/2007/11/010 
  [hep-th/0702165].
  %%CITATION = doi:10.1088/1475-7516/2007/11/010;%%
  
\bibitem{Durrer}
N.~Deruelle and V.~F.~Mukhanov,
  ``On matching conditions for cosmological perturbations,''
  Phys.\ Rev.\ D {\bf 52}, 5549 (1995)
  doi:10.1103/PhysRevD.52.5549
  [gr-qc/9503050];;\\
  %%CITATION = doi:10.1103/PhysRevD.52.5549;%%    
 R.~Durrer and F.~Vernizzi, 
``Adiabatic perturbations in pre - big bang models: Matching conditions and scale invariance,'' 
Phys.\ Rev.\ D {\bf 66}, 083503 (2002) 
doi:10.1103/PhysRevD.66.083503
[hep-ph/0203275];\\
%%CITATION = doi:10.1103/PhysRevD.66.083503;%%
C.~Cartier, R.~Durrer and E.~J.~Copeland,
  ``Cosmological perturbations and the transition from contraction to expansion,''
  Phys.\ Rev.\ D {\bf 67}, 103517 (2003)
  doi:10.1103/PhysRevD.67.103517
  [hep-th/0301198].
  %%CITATION = doi:10.1103/PhysRevD.67.103517;%%
   
\bibitem{NBV}
A.~Nayeri, R.~H.~Brandenberger and C.~Vafa,
  ``Producing a scale-invariant spectrum of perturbations in a Hagedorn phase of string cosmology,''
  Phys.\ Rev.\ Lett.\  {\bf 97}, 021302 (2006)
  doi:10.1103/PhysRevLett.97.021302
  [hep-th/0511140];\\
  %%CITATION = doi:10.1103/PhysRevLett.97.021302;%%
R.~H.~Brandenberger, A.~Nayeri, S.~P.~Patil and C.~Vafa,
  ``Tensor Modes from a Primordial Hagedorn Phase of String Cosmology,''
  Phys.\ Rev.\ Lett.\  {\bf 98}, 231302 (2007)
  doi:10.1103/PhysRevLett.98.231302
  [hep-th/0604126].
  %%CITATION = doi:10.1103/PhysRevLett.98.231302;%% 
  
\bibitem{BV}
R.~H.~Brandenberger and C.~Vafa,
 ``Superstrings In The Early Universe,'' 
 Nucl.\ Phys.\ B {\bf 316}, 391 (1989).
 %%CITATION = NUPHA,B316,391;%%  
 
\bibitem{Hagedorn}
R.~Hagedorn,
``Hadronic matter near the boiling point,''
Nuovo Cim. A \textbf{56}, 1027-1057 (1968)
doi:10.1007/BF02751614

\bibitem{GSW}
M.~B.~Green, J.~H.~Schwarz and E.~Witten,
  ``Superstring Theory. Vol. 1: Introduction,''
  Cambridge, Uk: Univ. Pr. ( 1987) 469 P. ( Cambridge Monographs On Mathematical Physics);\\
  M.~B.~Green, J.~H.~Schwarz and E.~Witten,
  ``Superstring Theory. Vol. 2: Loop Amplitudes, Anomalies And Phenomenology,''
  Cambridge, Uk: Univ. Pr. ( 1987) 596 P. ( Cambridge Monographs On Mathematical Physics).

%\cite{Brahma:2021tkh}
\bibitem{Brahma}
S.~Brahma, R.~Brandenberger and S.~Laliberte,
``Emergent Cosmology from Matrix Theory,''
[arXiv:2107.11512 [hep-th]].  

 \bibitem{Vafa}
 %\cite{Agrawal:2020xek}
P.~Agrawal, S.~Gukov, G.~Obied and C.~Vafa,
``Topological Gravity as the Early Phase of Our Universe,''
[arXiv:2009.10077 [hep-th]].

 
\end{thebibliography}
\end{document}